\documentclass[a4paper,10pt,twocolumn]{scrartcl} 

\usepackage{float}				
\usepackage{amsmath}			
\usepackage{amsfonts}			
\usepackage{amssymb}			
\usepackage{graphicx}			
\usepackage[T1]{fontenc}		
\usepackage{ae}					
\usepackage{typearea}			
\usepackage{scrpage2}			
\usepackage{lastpage}			
\usepackage[margin=10pt,font=small,labelfont=bf]{caption} 
\usepackage{textcomp}			
\usepackage{miller}             

\typearea{16}					
\begin{document}
\cfoot{\thepage /\pageref{LastPage}}     

\twocolumn[{\csname @twocolumnfalse\endcsname                

\title{\Large{Adsorption geometry and electronic structure of iron phthalocyanine on Ag surfaces: A LEED and photoelectron momentum mapping study}}
\author{\large{V. Feyer,$^{1,2,*}$ M. Graus,$^{3,4}$ P. Nigge,$^{3,4}$ M. Wie\ss{}ner,$^{3,4}$}\\
\large{R. G. Acres,$^2$ C. Wiemann,$^1$ C. M. Schneider,$^{1,5}$ A. Sch\"oll,$^{3,4}$ and F. Reinert,$^{3,4}$}}

\date{\small{\textit{
$^{1}$Peter Gr\"unberg Institute (PGI-6) and JARA-FIT, Research Center J\"ulich, 52425 J\"ulich, Germany.\\
$^{2}$Sincrotrone Trieste S.C.p.A., S.S. 14, km 163.5 in Area Science Park, 34012 Basovizza, Trieste, Italy\\
$^{3}$Universit\"at W\"urzburg, Experimentelle Physik VII \& R\"ontgen Research\\Center for Complex Material Systems RCCM, 97074 W\"urzburg, Germany\\
$^{4}$Karlsruher Institut f\"ur Technologie KIT, Gemeinschaftslabor f\"ur Nanoanalytik, 76021 Karlsruhe, Germany\\
$^{5}$Fakult\"at f. Physik and Center for Nanointegration Duisburg-Essen (CENIDE),\\ Universit\"at Duisburg-Essen, D-47048 Duisburg, Germany}}\\
\hspace{1ex}
\\$^*$Corresponding author: v.feyer@fz-juelich.de\\(Dated: September 13, 2013)}

\maketitle                                                      
\begin{abstract}                                                
We present a comprehensive study of the adsorption behavior of iron phthalocyanine on the low-index crystal faces of silver. By combining measurements of the reciprocal space by means of photoelectron momentum mapping and low energy electron diffraction, the real space adsorption geometries are reconstructed. At monolayer coverage ordered superstructures exist on all studied surfaces containing one molecule in the unit cell in case of Ag(100) and Ag(111), and two molecules per unit cell for Ag(110). The azimuthal tilt angle of the molecules against the high symmetry directions of the substrate is derived from the photoelectron  momentum maps. A comparative analysis of the  momentum patterns on the substrates with different symmetry indicates that both constituents of the twofold degenerate FePc lowest unoccupied molecular orbital are occupied by charge transfer from the substrate at the interface.
\\
\\ 
\end{abstract}
}]

\section{Introduction}
Organometallic oligomers such as phthalocyanines (Pc) and porphyrins deposited on a noble metal surface have a tendency to self organize and form highly ordered patterns, making them attractive for technological applications in colorimetric gas sensors \cite{2}, \cite{3}, photonic wires, field-effect transistors, light emitting diodes, catalysts, optical switches \cite{4}, \cite{5} and data storage \cite{6}. Recently, it was shown that organic semiconductors have a great opportunity for application in the growing area of spintronics, where active manipulation and control of the electronic spin degree of freedom are of interest \cite{7}, \cite{8}. 

Most of the metal-Pcs are planar molecules formed by stable $\pi$-conjugated macromolecules coordinating to a central metal atom whereby appropriate selection of the metal can tune the electronic structure of the attendant Pc. However, the geometry can be modified to non planar configurations by placing e.\,g. Sn or TiO in the central part of Pc molecules \cite{9}.

The organic-inorganic interface plays a crucial role in understanding both the morphology and the electronic properties of epitaxial thin films. In particular, the lowest unoccupied and highest occupied molecular orbitals (LUMO and HOMO, respectively) are of interest since they determine the bonding mechanisms of molecules on surfaces \cite{10}, \cite{11}, \cite{Ziroff-PRL2010}.

In this work we present a comprehensive study of the adsorption behavior of FePc at a saturated ($\sim$\,1\,monolayer\,=\,1\,ML) coverage on the low-index crystal faces of silver. The (110), (100) and (111) orientations were chosen for a systematic variation of the surface symmetry (i.e. twofold, fourfold and sixfold, respectively). In consequence, a different number of symmetry-equivalent superstructure domains can be expected on these surfaces. We have used a combination of LEED and photoelectron momentum mapping (PMM) that provides access to the superstructure unit cell and the arrangement of the molecules within the unit cell \cite{wiessner}, \cite{willenbockel}. Additionally momentum patterns can give information about the electronic structure at the interface and the symmetry of molecular orbitals \cite{puschnig}. 

\begin{figure*}[t]		
\begin{center}
\begin{minipage}[t]{0.17\textwidth}
\textbf{FePc on:}\\ \vspace{11ex}\\ 
\textbf{submonolayer}\\ \vspace{19ex}\\ 
\textbf{monolayer}\\ \vspace{20ex}\\ 
\end{minipage}
\begin{minipage}[t]{0.27\textwidth}
\begin{center}
\textbf{Ag(100)}\\ \hspace{2ex}\\
\includegraphics[width=0.9\linewidth]{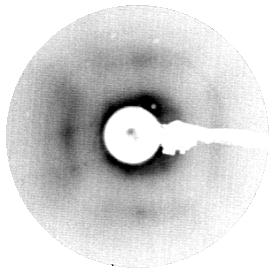}   
\includegraphics[width=0.9\linewidth]{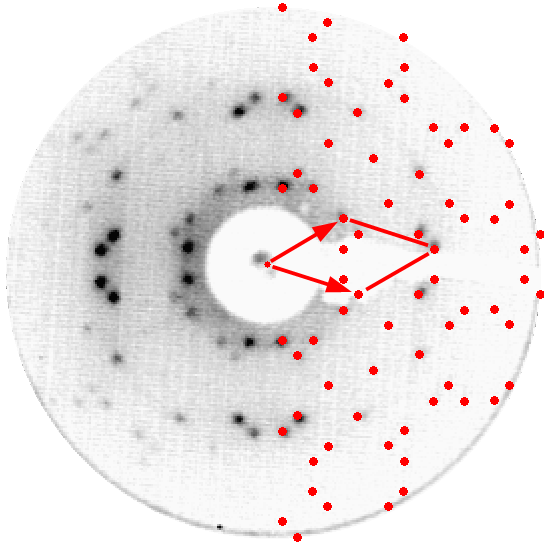}   
\end{center}                                   
\end{minipage}
\begin{minipage}[t]{0.27\textwidth}
\begin{center}
\textbf{Ag(110)}\\ \hspace{2ex}\\
\includegraphics[width=0.9\linewidth]{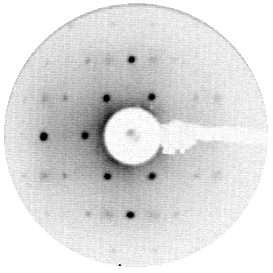}   
\includegraphics[width=0.9\linewidth]{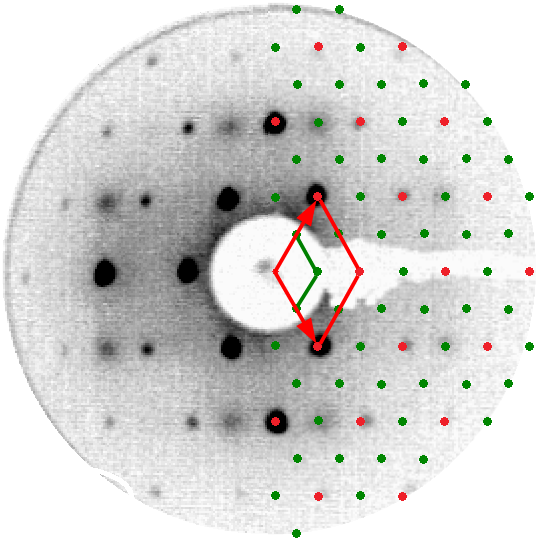} 
\end{center}                
\end{minipage}
\begin{minipage}[t]{0.27\textwidth}
\begin{center}
\textbf{Ag(111)}\\ \hspace{2ex}\\
\includegraphics[width=0.9\linewidth]{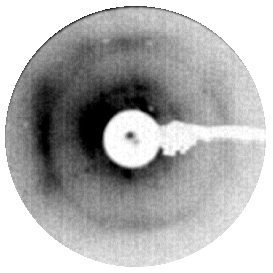}   
\includegraphics[width=0.9\linewidth]{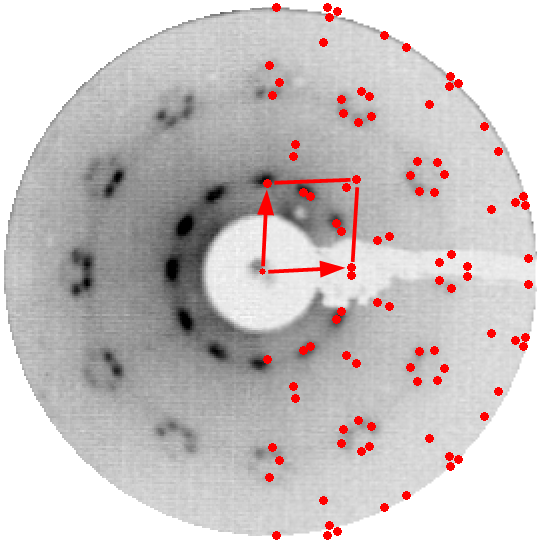} 
\end{center}                
\end{minipage}
\vspace{-10ex}
\caption[]{(color online.) LEED patterns of FePc adsorbed on Ag(100), Ag(110) and Ag(111) at submonolayer (bottom) and saturated monolayer coverage (top) (beam energy 14 eV). The corresponding superstructure simulations and reciprocal unit cells are displayed by red dots and arrows together with the monolayer patterns. For FePc/Ag(110) the red dots/arrows represent the c(10\,$\times$\,4) superstructure whereas green dots/arrows belong to a p(10\,$\times$\,4) superstructure.}
\label{Fig1} 
\end{center}
\end{figure*}

\section{Experimental details}
Photoelectron momentum microscopy and LEED experiments were performed at the NanoESCA beamline of the Elettra synchrotron \cite{20} in an ultrahigh vacuum (UHV) system with a base pressure of 2$\times$10$^{-10}$\,mbar. PMM was performed using the NanoESCA photoemission spectrometer. The set-up includes a non-magnetic, electrostatic photoelectron emission microscope (PEEM) and a double-pass hemispherical analyser. This instrument is equipped with a transfer lens behind the immersion lens objective to map the angular distribution of the photoelectrons by imaging the focal plane. In k-imaging mode, the microscope can detect energy resolved photoemission intensities in the whole emission hemisphere above the sample surface with extreme efficiency \cite{30}. The energy resolution was estimated to 150\,meV and the momentum resolution as $\pm$\,0.05\,\r{A}$^{-1}$, using p-linear polarized light. 

For LEED a SPECS LEED system with an attached CCD camera and the SAFIRE software was utilized. 

Clean surfaces of Ag(111), Ag(110) and Ag(100) were prepared by standard procedures of cycles of Ar+ ion sputtering (kinetic energy 2.0\,keV), followed by annealing at 800\,K. The surface order and cleanliness were monitored by LEED and PES.

For film preparation a few mg of FePc (Alfa Aesar GmbH, 96\,\% purity) were loaded into home-made Knudsen cell type evaporators connected to separate preparation chambers. The FePc was carefully degassed at 600\,K for several days while monitoring the base pressure of the UHV system. The molecules were thermally evaporated onto the Ag substrates at room temperature. After deposition all samples were annealed to 400\,K in order to allow diffusion of molecules over the step edges and to obtain uniform coverage on all surface terraces. 

All experiments were done at room temperature. 

For the simulation of the momentum pattern the FePc HOMO and LUMO were calculated in analogy to Ref. \cite{Marom-APA2009}. 

\section{Results and discussion}
\subsection{Lateral order}
The LEED patterns of low (submonolayer >\,0.5\,ML) and saturated coverage ($\sim$\,1\,ML) of FePc on the (100), (110) and (111) Ag surfaces are shown in Fig\,\ref{Fig1}. At low coverage, the diffraction patterns show a ring-like superstructure on Ag(100) and Ag(111) which corresponds to the formation of a two dimensional gas phase (g-phase). This is in agreement with recent STM studies \cite{21}. Similar g-phase patterns were also observed for low coverages of H$_2$Pc, Fe and CuPc on Ag(111) \cite{21}, \cite{16}, \cite{22}. In contrast, FePc/Ag(110) shows a sharp LEED pattern at low coverage with a c(10$\times$4) superstructure with the matrix \hbox{(5, 2 / 5, -2)} which is conserved upon increasing the coverage to 1\,ML and is in agreement with recent STM studies \cite{15}, which also demonstrated that the FePc molecules in the first ML lie flat on the surface. It was found that the molecular symmetry axis bisecting the FePc phenyl rings is oriented at an angle of either 45\textdegree\,(with a corresponding c(10\,$\times$\,4) phase) or $\pm$\,30\textdegree\,(with two alternating p(10\,$\times$\,4) structures)  with respect to the \hkl[-1 1 0] direction of the substrate \cite{15}. While it was shown by STM that both phases coexist, the LEED pattern of FePc/Ag(110) showed only a c(10\,$\times$\,4) superstructure in \cite{15}. This was attributed to the fact that the LEED pattern is mainly determined by scattering from the heavy Fe atoms, that form a c(10\,$\times$\,4) subnet in both superstructures \cite{15}. Our high-contrast LEED data in Fig.\,\ref{Fig1} provide additional information and can help to solve this inconsistency. In addition to the bright spots associated to a c(10\,$\times$\,4) superstructure some weaker spots can be detected, which belong to a p(10\,$\times$\,4) LEED pattern. We can hereby conclude, that both reported FePc phases may coexist on the sample according to LEED. The length of the corresponding unit cell basis vector in real space is approximately 17\,\r{A} for the c(10\,$\times$\,4) and 33\,\r{A} for the p(10\,$\times$\,4) superstructure, which allows one and two FePc molecules (sidelength $\sim$\,12\,\r{A}) per unit cell, respectively.   

The LEED pattern for a monolayer FePc/Ag(100) shows a superlattice matrix of \hbox{(5, 1 / 2, 6)}, which refers to a unit cell with real space basis vector lengths of 15\,\r{A} and 18\,\r{A}. The saturated coverage of FePc on the Ag(111) surface forms an incommensurate superstructure (see Fig.\,\ref{Fig1}) with a matrix of (4.65, -0.3 / 2.55, 5.57) which implies a length of the molecular superlattice basis vectors of 13.9\,\r{A} and 14.0\,\r{A} and an enclosed angle of 96\,\textdegree. These values are in good agreement with previously published results \cite{21}. 

In the course of the following analysis it will be of importance that the superstructure unit cells determined for FePc/Ag(100) and FePc/Ag(111) by LEED are too small to contain more than one flat lying FePc molecule.  

\subsection{Electronic structure and azimuthal orientation}

\begin{figure}
\begin{center}
\includegraphics[width=0.9\linewidth]{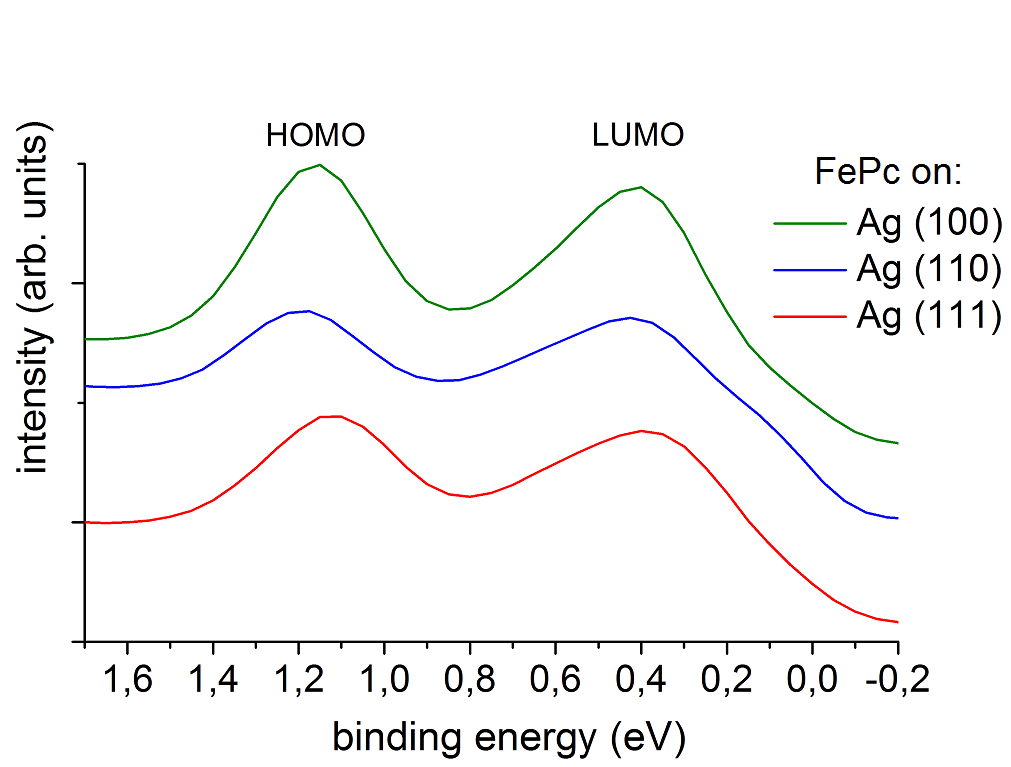}
\caption[]{(color online.) PES EDCs of 1\,ML FePc on different Ag surfaces, recorded with the NanoESCA at a photon energy of 26\,eV at room temperature by integration over an area of about 1/2 \r{A}$^{-1}$ diameter with comparable HOMO and LUMO signal. } 
\label{Fig2}  
\end{center}
\end{figure}

Fig.\,\ref{Fig2} shows the PES energy distribution curves (EDCs) of 1\,ML FePc on Ag(100), Ag(110) and Ag(111). The EDCs were derived by integrating over a part of the photoemission hemisphere at $k_x$\,=\,0.5\,\AA$^{-1}$, $k_y$\,=\,1.5\,\AA$^{-1}$ with a diameter of about 1/2 \r{A}$^{-1}$, which leads to a slightly better energy resolution and calibration as compared with integration over the whole hemisphere.
 The energy was calibrated by fitting a Fermi-function to the low-binding energy onset, resulting in an accuracy of the energy scale of about 0.1\,eV. The asymmetric peak closest to the Fermi edge at 0.4\,eV binding energy is assigned to the partially filled FePc LUMO, which is occupied by charge transfer from the substrate. At a binding energy of approximately 1.2\,eV the FePc HOMO is observed. The energy positions of the HOMO and LUMO are consistent with previously published PES data \cite{12}. 

\begin{figure*}[t]		
\begin{center}
\begin{minipage}[t]{0.32\textwidth}
\begin{center}
\textbf{FePc on:\hspace{4ex} Ag(100)}\hfill\hbox{ }\\ \hspace{2ex}\\
\includegraphics[width=1\linewidth]{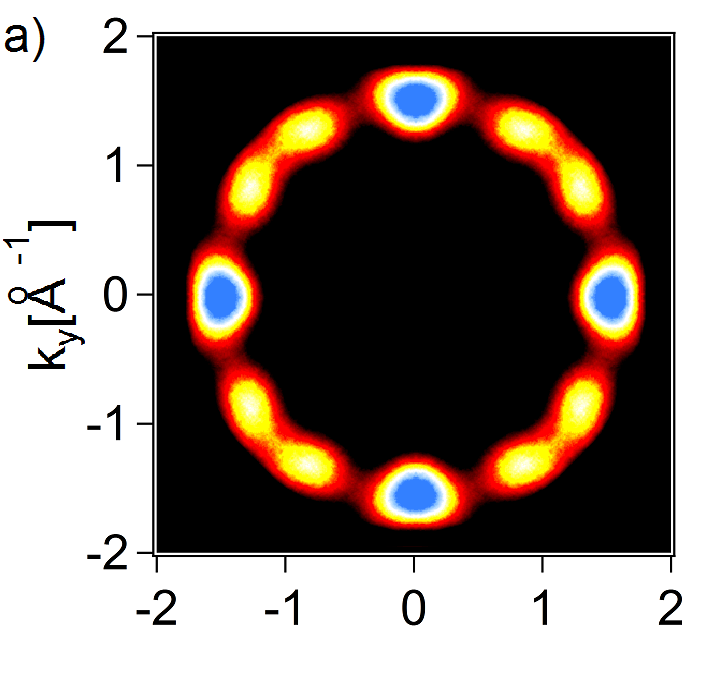}
\includegraphics[width=1\linewidth]{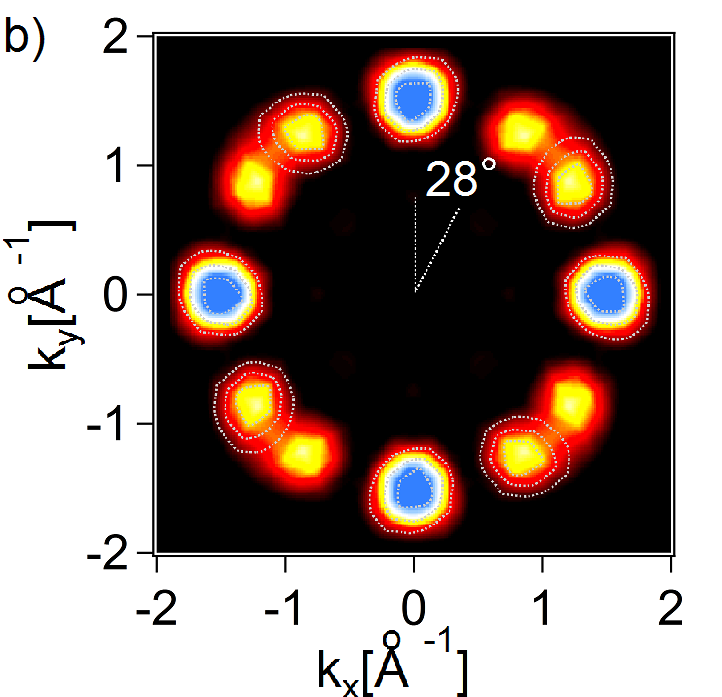}\vspace{5ex}
\includegraphics[width=1\linewidth]{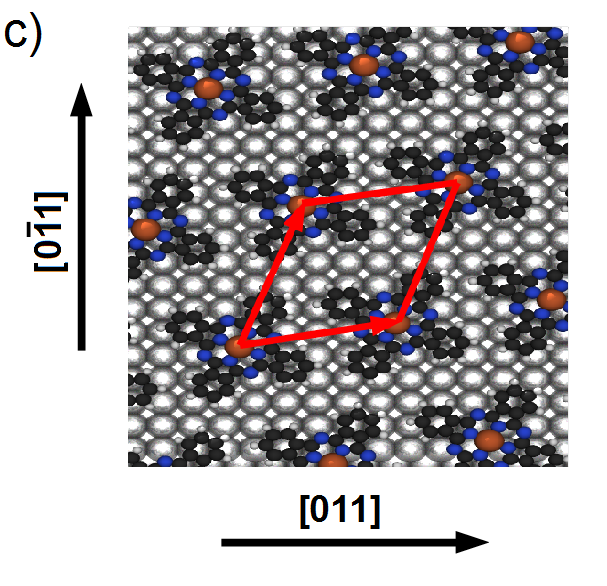} 
\end{center}                                   
\end{minipage}
\begin{minipage}[t]{0.32\textwidth}
\begin{center}
\textbf{\hspace{4ex}Ag(110)}\\ \hspace{2ex}\\
\includegraphics[width=1\linewidth]{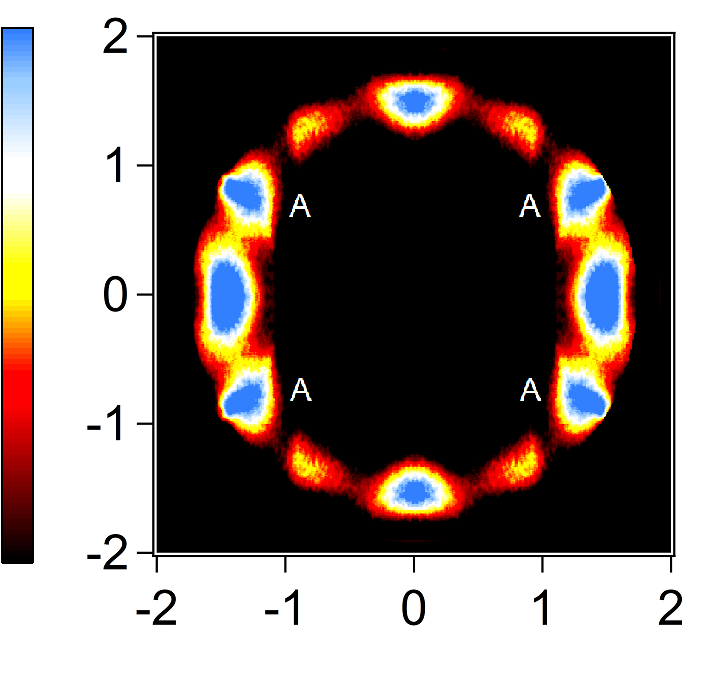}
\includegraphics[width=1\linewidth]{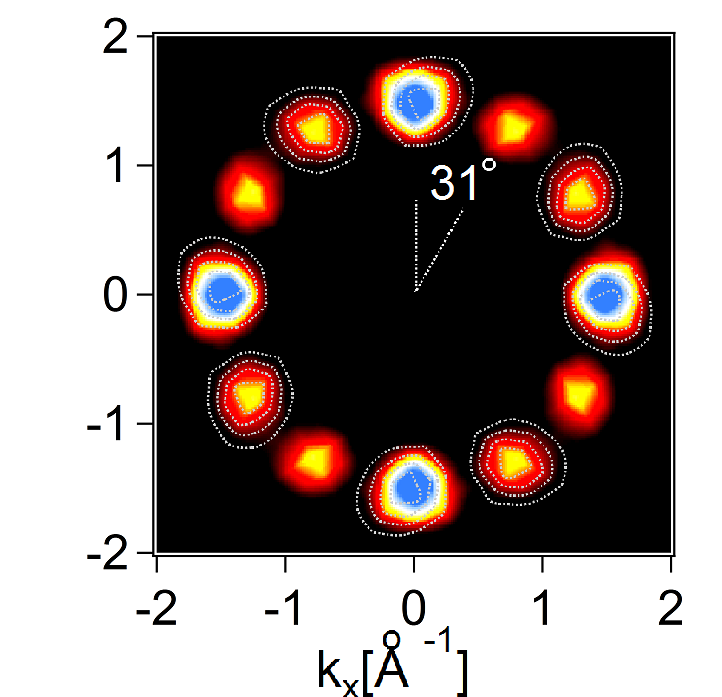}\vspace{5ex}
\includegraphics[width=1\linewidth]{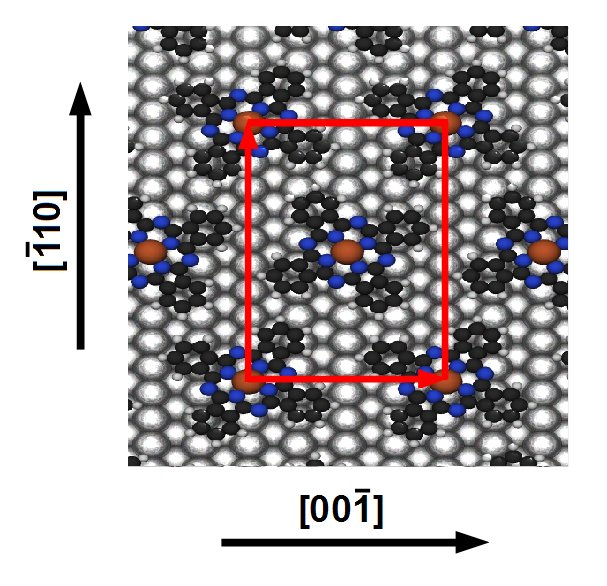}
\end{center}                
\end{minipage}
\begin{minipage}[t]{0.32\textwidth}
\begin{center}
\textbf{\hspace{4ex}Ag(111)}\\ \hspace{2ex}\\
\includegraphics[width=1\linewidth]{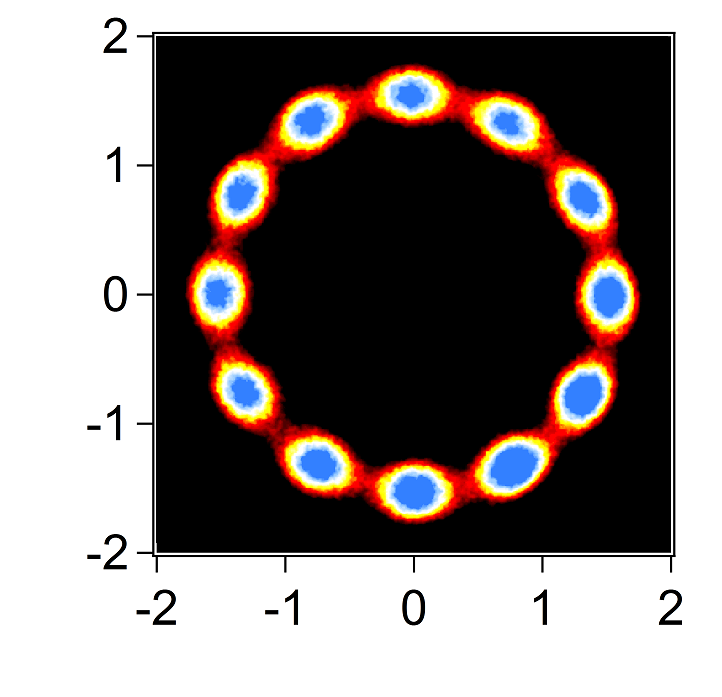}   
\includegraphics[width=1\linewidth]{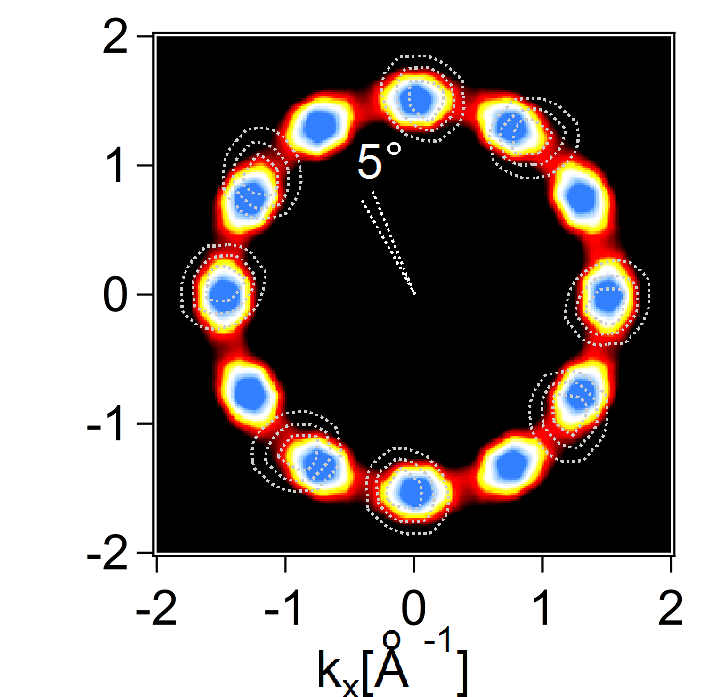}\vspace{5ex}
\includegraphics[width=1\linewidth]{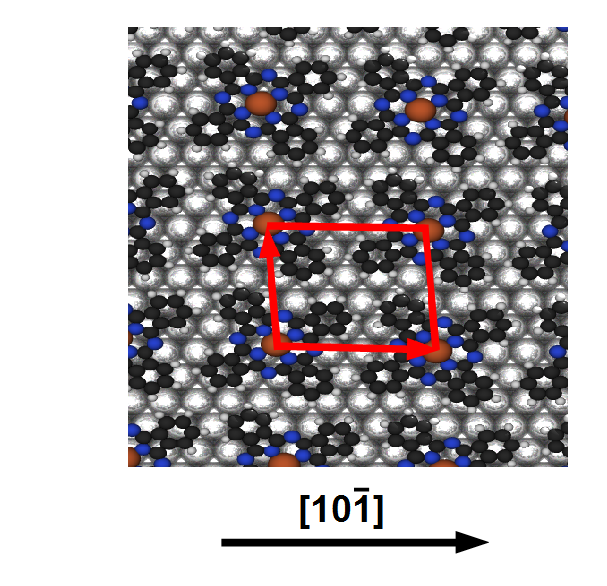}
\end{center}                
\end{minipage}
\caption[]{(color online.)\\
a) Experimental photoelectron momentum maps (PMM) determined for photoelectrons from the FePc HOMO region (50\,meV energy window around 1.2\,eV) for 1\,ML FePc on Ag(100), Ag(110) and Ag(111). The incoming light was p-polarized with a photon energy of 26\,eV. The data was symmetrized according to the substrate symmetry (i.e. fourfold, twofold and sixfold, respectively), which suppresses the $\vec{A}$\,$\cdot$\,$\vec{k}$ dependency introduced by the geometry of the experiment. In case of Ag(110) the intense signal from the sp band contributes to the features marked by A.\\
b) Simulated PMM for the FePc HOMO at a kinetic energy of 20\,eV for a superposition of different metal-Pc domains azimuthally rotated by $\pm$\,28\textdegree, $\pm$\,31\textdegree\, and $\pm$\,5\textdegree\,, respectively, against the high symmetry directions (see text for details). Dashed lines show the intensity distributions of a single domain.\\
c) Illustration of the real space adsorption geometry derived from LEED and PMM. The adsorption site was chosen arbitrarily. For Ag(100) and Ag(111), in each case only one of the existent domains is shown. For Ag(110), the pattern with alternating rotated molecules ($\pm$\,31\textdegree) is depicted.}
\label{Fig3} 
\end{center}
\end{figure*}

For further analysis of the charge distribution in the HOMO and LUMO of the FePc/Ag system and for the investigation of the molecular orientation, we have utilized the NanoESCA to perform PMM measurements of the respective samples. It has been demonstrated recently that the measured momentum  patterns can be interpreted as the square of the Fourier transforms of the wavefunction in real space, thus allowing identifying corresponding molecular orbitals \cite{puschnig}, \cite{Ziroff-PRL2010}, \cite{dauth-prl2011}, \cite{Puschnig-PRB2012} and the azimuthal alignment of adsorbed molecules \cite{wiessner}, \cite{willenbockel}. The measured angle dependent photoelectron intensity distributions derived from a 50\,meV energy window in the HOMO region for 1\,ML FePc on the different Ag substrates are shown in Fig.\,\ref{Fig3}\,a. The two-dimensional intensity patterns were symmetrized according to the respective substrate symmetry. By this evaluation also the $\vec{A}$\,$\cdot$\,$\vec{k}$ dependency introduced by the geometry of the experiment can be suppressed. The resulting momentum maps feature several characteristic intensity maxima.

According to Ref. \cite{puschnig} the momentum maps can be simulated by Fourier transformation of the respective real space orbitals at the corresponding kinetic energy of the photoelectrons of about 20\,eV.

\begin{figure*}[t]		
\begin{center}
\begin{minipage}[t]{0.32\textwidth}
\begin{center}
\textbf{FePc on:\hspace{4ex} Ag(100)}\hfill\hbox{ }\\ \hspace{2ex}\\
\includegraphics[width=1\linewidth]{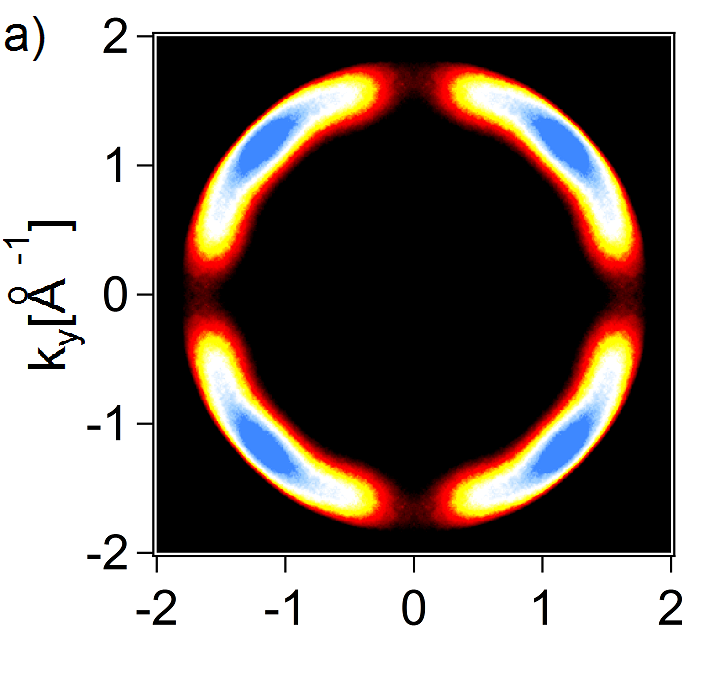}
\includegraphics[width=1\linewidth]{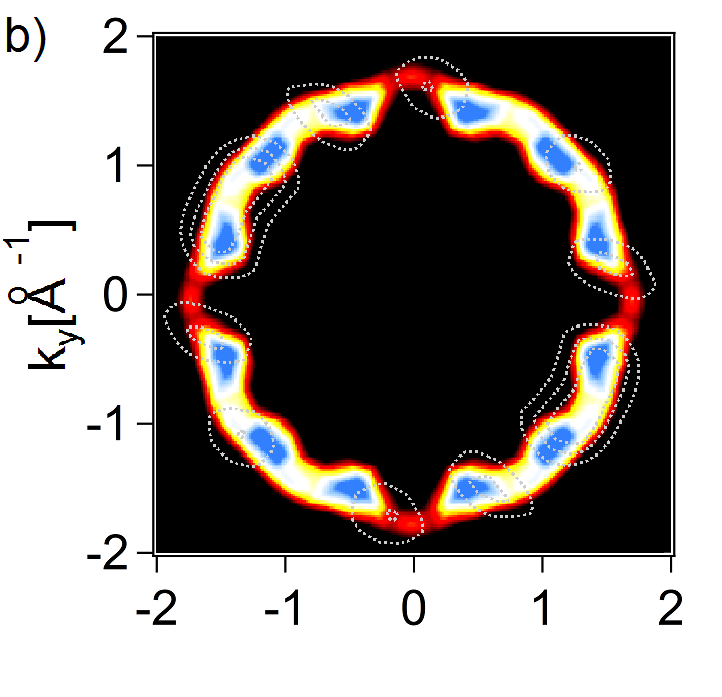}
\includegraphics[width=1\linewidth]{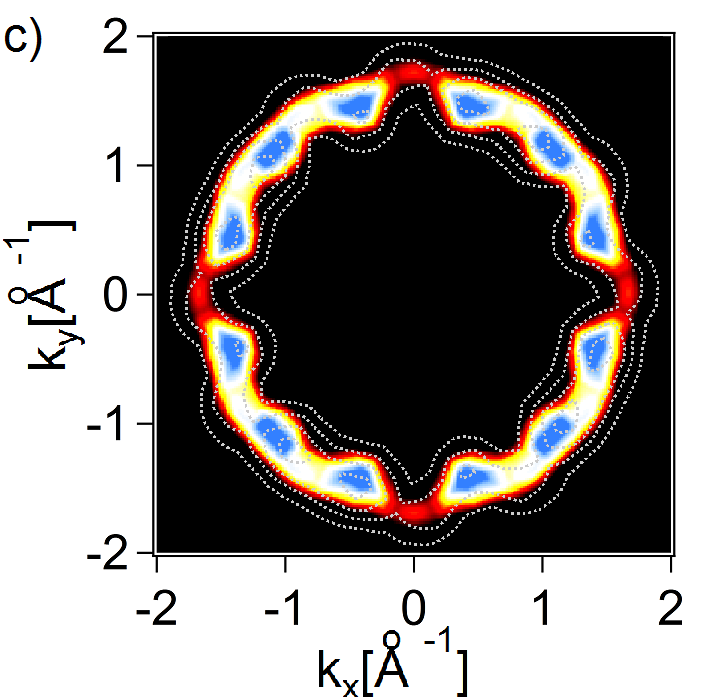} 
\end{center}                                   
\end{minipage}
\begin{minipage}[t]{0.32\textwidth}
\begin{center}
\textbf{\hspace{4ex}Ag(110)}\\ \hspace{2ex}\\
\includegraphics[width=1\linewidth]{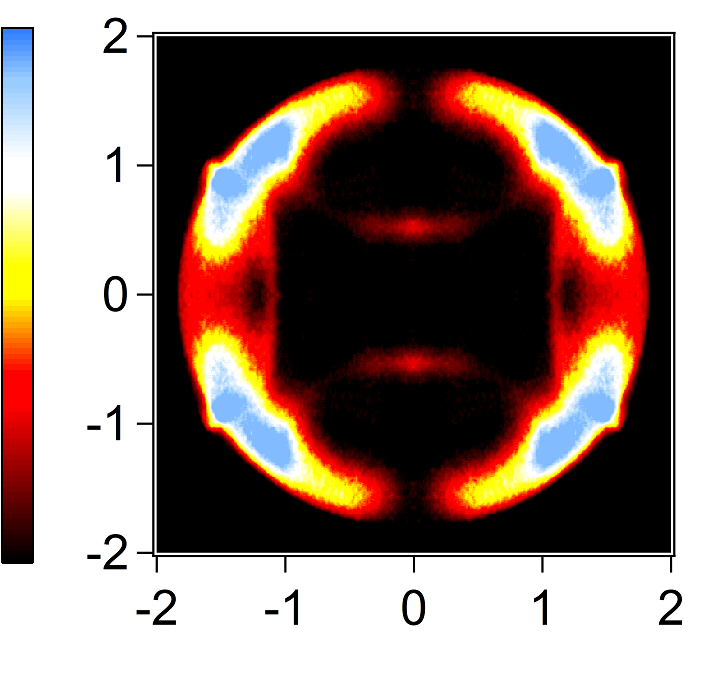}
\includegraphics[width=1\linewidth]{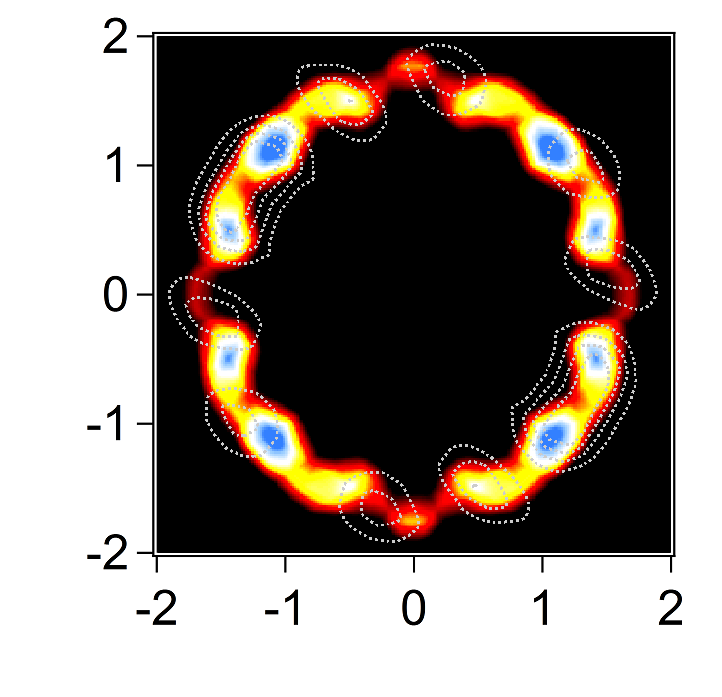}
\includegraphics[width=1\linewidth]{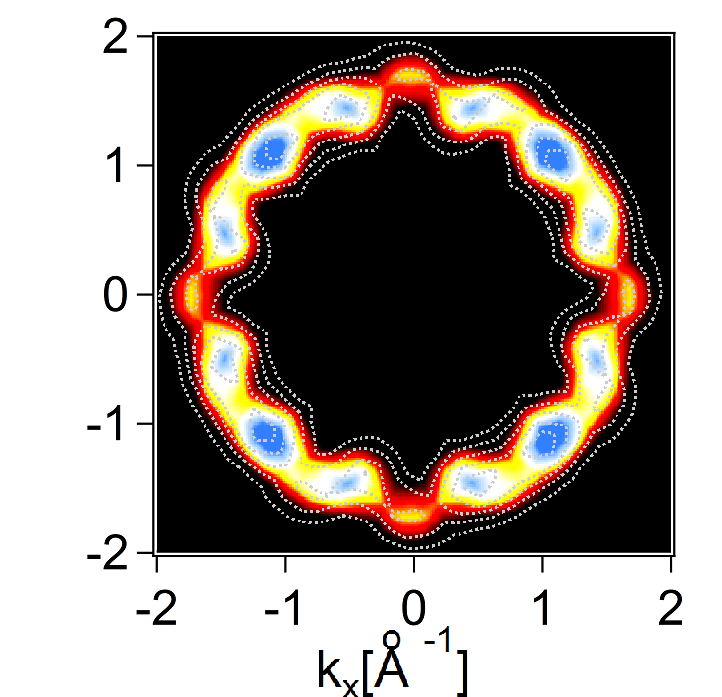}
\end{center}                
\end{minipage}
\begin{minipage}[t]{0.32\textwidth}
\begin{center}
\textbf{\hspace{4ex}Ag(111)}\\ \hspace{2ex}\\
\includegraphics[width=1\linewidth]{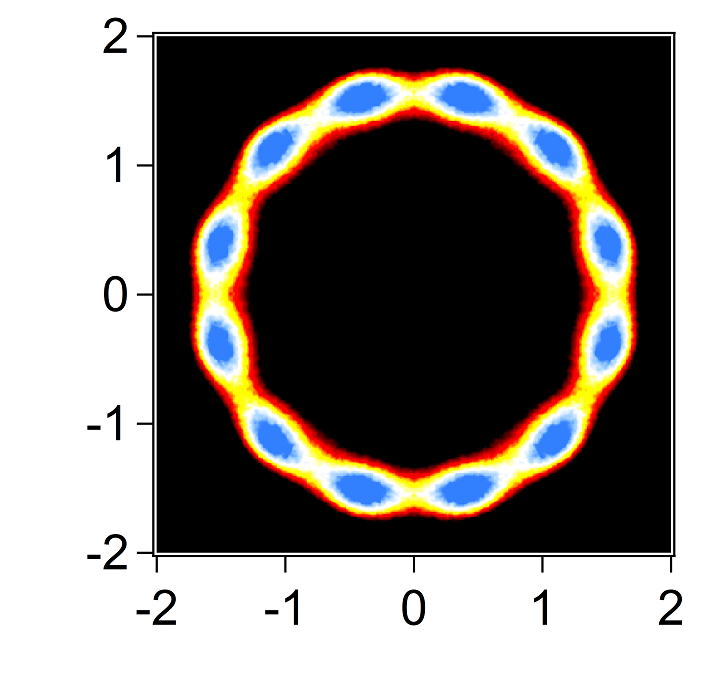}   
\includegraphics[width=1\linewidth]{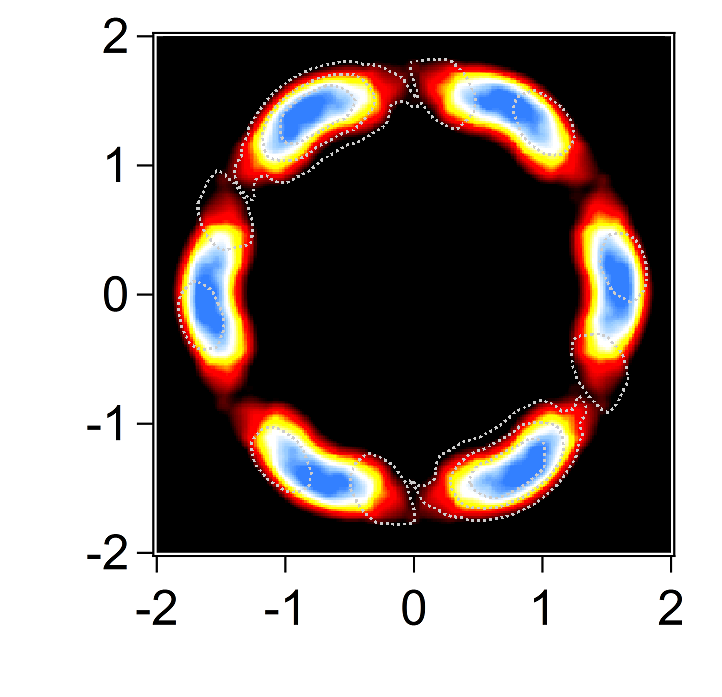}
\includegraphics[width=1\linewidth]{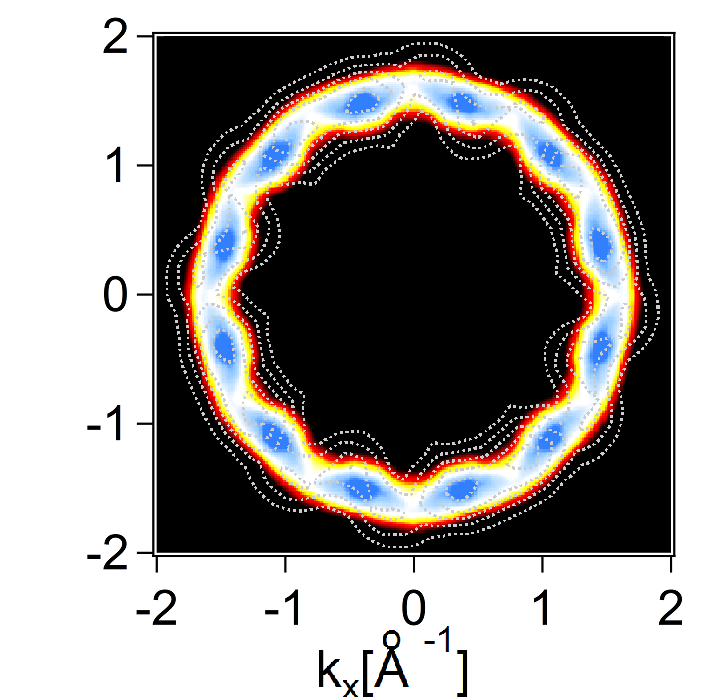}
\end{center}                
\end{minipage}
\caption[]{(color online.)\\
a) Experimental PMM derived for photoelectrons from the FePc LUMO region (50\,meV window around 0.4\,eV) for 1\,ML FePc on Ag(100), Ag(110) and Ag(111). The incoming light was p-polarized with a photon energy of 26\,eV.\\
b) Simulated PMM for the FePc LUMO at a kinetic energy of 20\,eV for the azimuthal arrangement determined from the FePc HOMO (see  Fig.\,\ref{Fig3}). Only one of the two originally degenerate LUMOs was considered. The dashed lines show the signal from only one domain.\\
c) The same simulations as in b), but considering the signal from both degenerate LUMOs.}
\label{Fig4} 
\end{center}
\end{figure*}

\clearpage

 From a comparison of the theoretical momentum maps with experiment it becomes immediately clear that the FePc molecules must be adsorbed with a certain azimuthal tilt against the substrates' high symmetry directions. Since we furthermore know from our LEED analysis, delineated above, that the superstructure unit cell contains one molecule for FePc/Ag(100), the symmetry of the substrate will lead to two mirror domains with opposite tilt of the FePc molecules. In case of FePc/Ag(100) the simulation matches the experiment best for tilt angles of $\pm$\,28\textdegree\, against the  \hkl[0 -1 1]-direction. For FePc/Ag(110) two FePc molecules with tilt angles of $\pm$\,31\textdegree\, against \hkl[-1 1 0] have to be considered.  For FePc/Ag(111) a tilt of $\pm$\,5\textdegree\ against the \hkl[-1 1 0]-direction with consideration of the sixfold surface symmetry provided the best simulation. The respective images are displayed in Fig.\,\ref{Fig3}b.
 
\subsection{Real space models}
Since the size of the superstructure unit cell requires one molecule per unit cell the models for the arrangement of the FePc molecules on the Ag(100) and on the Ag(111) surface can be straightforwardly derived from the knowledge of the unit cell and the tilt angle of the molecules against the substrates' high symmetry direction. The exact adsorption site, however, cannot be determined in our experiments. Illustrations of the respective real space models are displayed in Fig.\,\ref{Fig3}c with arbitrarily chosen adsorption sites. In case of FePc/Ag(110) the situation is more complicated. Two possible scenarios may occur. For the c(10\,$\times$\,4) superstructure one molecule per unit cell with a tilt of \,31\textdegree\,against \hkl[1 -1 0] (and the corresponding mirror domain) can be concluded. For the p(10\,$\times$\,4) two molecules per unit cell must occur with tilt angles of $\pm$\,31\textdegree\,. Our data does not prefer one of these possibilities. However, a comparison to literature shows, that our proposed  p(10\,$\times$\,4) phase matches STM results nicely, while in the same experiments a 45\textdegree\, tilt of the FePc molecules in the c(10\,$\times$\,4) phase \cite{15} was found. The latter is incompatible with our PMM results and thus rendering the p(10\,$\times$\,4) more likely. The FePc/Ag(110) real space model in Fig.\,\ref{Fig3}c thus displays this phase.  

\subsection{LUMO momentum maps}
Fig.\,4\,a shows the momentum maps derived for the FePc LUMO on the three different substrates. As already seen for the HOMO on Ag(110), the LUMO momentum pattern for this surface direction is influenced by the Ag sp band leading to some spurious background. It is an interesting issue that the LUMO of many Pcs is twofold degenerate. This obviously also occurs for FePc, as derived from calculations \cite{Marom-APA2009}. Marom et al. moreover find a Jahn-Teller splitting that lifts the degeneracy by about 200\,meV in the gas phase \cite{Marom-APA2009}. In case of adsorption this splitting may be increased due to the symmetry mismatch and the distortion following the interfacial bonding \cite{Stadler-NatPhys}. It is thus not clear a priori, whether the charge transfer into the molecule, which is evident from our PES in Fig.\,\ref{Fig2}, involves both LUMOs or only one of them. The latter scenario has been proposed for CuPc/Ag(111) \cite{Stadtmueller}, were the authors respect only one of the degenerate LUMOs for the simulation of their angular resolved photoemission  data. In order to test these possibilities we have simulated the LUMO momentum maps considering one and two LUMOs. The corresponding images are displayed in Fig.\,4\,b and c, respectively. The azimuthal tilt angles are already known from the analysis of the HOMO.  

Since the two degenerate LUMOs of FePc enclose an angle of 90\,\textdegree, for Ag(100) with its 90\,\textdegree\,invariance superpositions of different domains based on a single LUMO (see Fig.\,4\,b) show the same pattern as those based on both LUMOs (Fig.\,4\,c). This is different for the Ag(110) surface, since it is only twofold symmetric, though the differences in the corresponding images in Figs.\,4\,b and c are small and insignificant due to the silver band background. For Ag(111) however, significant differences occur between the two scenarios. The comparison of the simulations respecting one and two degenerate LUMOs with the experimental momentum map shows, that both LUMOs have to be taken into account. Charge transfer from the substrate thus obviously occurs into both LUMOs.

\section{Conclusion}
We have analysed the geometric structure of FePc on Ag(100), Ag(110) and Ag(111) substrates. Combining LEED and PMM we have derived the superstructure unit cells and the arrangement of the molecules within the unit cell. While for FePc/Ag(100) and FePc/Ag(111) the data is unambiguous and shows unit cells with one molecule that is azimuthally tilted by \,28\textdegree\, and \,5\textdegree\,, respectively, against the substrates high-symmetry directions, for FePc/Ag(110) a tilt angle of 31\textdegree\,is derived which is compatible with two molecules in a p(10\,$\times$\,4) unit cell and one molecule in a c(10\,$\times$\,4) unit cell. However, only the former scenario is consistent with literature \cite{15}. Moreover, the comparative analysis of the momentum maps of the LUMO of FePc on the substrates with different symmetry indicates that both degenerate LUMOs receive charge transfer from the substrate and are thus occupied at the interface.

\section{Acknowledgement}
A.S. and F.R. thank the Deutsche Forschungsgemeinschaft (grants GRK 1221 and RE1469/9-1) and the Bundesministerium f\"ur Bildung und Forschung (contract 03SF0356B) for financial support.

\end{document}